\documentclass[aps,amsmath,amssymb,prl,twocolumn,groupedaddress,showpacs]{revtex4}
\usepackage{bm}
\usepackage{epsfig}
\usepackage[usenames]{color}

\def\lsim{\mathrel{\rlap{\lower4pt\hbox{\hskip1pt$\sim$}}
    \raise1pt\hbox{$<$}}}                
\def\gsim{\mathrel{\rlap{\lower4pt\hbox{\hskip1pt$\sim$}}
    \raise1pt\hbox{$>$}}}                

\begin{document}

\title{  Signatures of neutral quantum Hall modes in transport through low-density 
constrictions   }

\author{Bernd Rosenow}
\affiliation{Physics Department, Harvard University, Cambridge, MA
02138, USA}
\author{Bertrand I.~Halperin}
\affiliation{Physics Department, Harvard University, Cambridge, MA
02138, USA} 
\date{June 11, 2008}

\begin{abstract}

Constrictions in  fractional quantum Hall (FQH) systems not only facilitate  backscattering 
between counter-propagating edge modes, but also may reduce  
 the  constriction filling fraction $\nu_c$ with respect to  the bulk filling fraction $\nu_b$. If both $\nu_b$ and $\nu_c$  correspond to incompressible FQH states, at least part of the constriction region is surrounded by  composite edges, 
whose low energy dynamics is characterized by a charge mode and one or several neutral modes. In the incoherent regime, decay of neutral modes describes the equilibration of composite FQH edges, while in the limit of coherent transport,
the presence of neutral modes gives rise to universal conductance
fluctuations. In addition, neutral modes
renormalize the strength of scattering across the constriction, and
thus can determine the relative strength of forward and backwards
scattering.
\end{abstract}

\pacs{73.43.-f,73.43.Cd, 73.43.Jn}

\maketitle

The strongly correlated nature of fractional quantum Hall (FQH) states 
is reflected in their unusual  low energy edge excitations.
Edges of simple FQH states realize  
  a chiral Luttinger liquid (LL)  \cite{Wen91}, and
backscattering 
between counter-propagating  FQH edges can be used to study their  dynamics.
 For FQH  quasiparticles, the backscattering amplitude is then expected to increase with decreasing source-drain voltage, giving rise to a zero bias peak in the differential resistance.

Scattering between FQH edges is facilitated 
 by a   constriction region,  where two counter-propagating edges approach each other closely. Due to the confining potential, the constriction filling fraction 
 $\nu_c$  is generally smaller than the bulk filling fraction $\nu_b$, and 
 interesting transport characteristics result. For $\nu_b=1$, a zero bias peak in 
 the differential resistance was experimentally observed for  $\nu_c < 1/2$, while for $\nu_c > 1/2 $ a zero bias dip was seen 
 instead  \cite{Rodarro+04,Roddaro+05} and interpreted in terms of 
 particle-hole transformations. \cite{Roddaro+05,Lal07}

In this letter, we analyze charge transport through a low density FQH constriction. 
If both  $\nu_b$ and  $\nu_c$ are incompressible states, the constriction 
region is surrounded by two types of edges: between $\nu_c$ and vacuum,
and between $\nu_b$ and $\nu_c$.  At least one of these edges is a 
composite edge with  counter-propagating   modes. 
If spatially random intra-edge scattering  is relevant, the low energy  physics of composite edges is  described by a random fixed point with  
a charged mode  decoupled from one or several neutral modes  
\cite{KaFiPo94,KaFi95}. Edge equilibration is described by the decay of 
the neutral mode, with  a  disorder-induced equilibration length $\ell$. 
Denoting the geometric constriction size by $L$ and the neutral mode velocity by $v_\sigma$, there are additional  length scales  $L_T= v_\sigma \hbar /k_B T$ determined by temperature and 
$L_V = v_\sigma \hbar / e V$ determined by the source-drain voltage. 
In the diffusive limit $\ell \ll L$ and ${\rm min}(L_T, L_V) < L$, 
transport through the constriction acquires a temperature and voltage
dependence through the neutral mode decay length $\ell(T,V)$. 
In the coherent limit $L_T, L_V > L$, $\ell < L$,  neutral edge modes have a 
dramatic influence on transport as they give rise to universal conductance 
oscillations. We find that the  conductance through the constriction can 
take any value  between zero and $\nu_b e^2/h$, and changes as a function of 
chemical potential or magnetic field.  

\begin{figure}[h]
\includegraphics[width=0.85\linewidth]{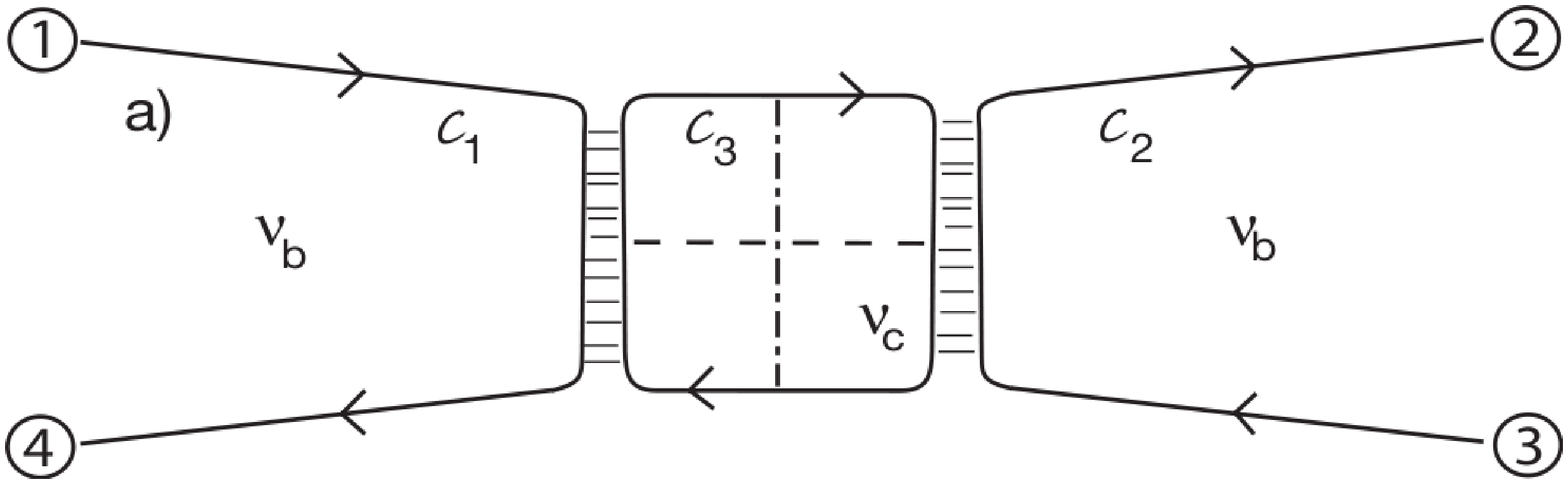}
\vspace*{.5cm}

\includegraphics[width=0.85\linewidth]{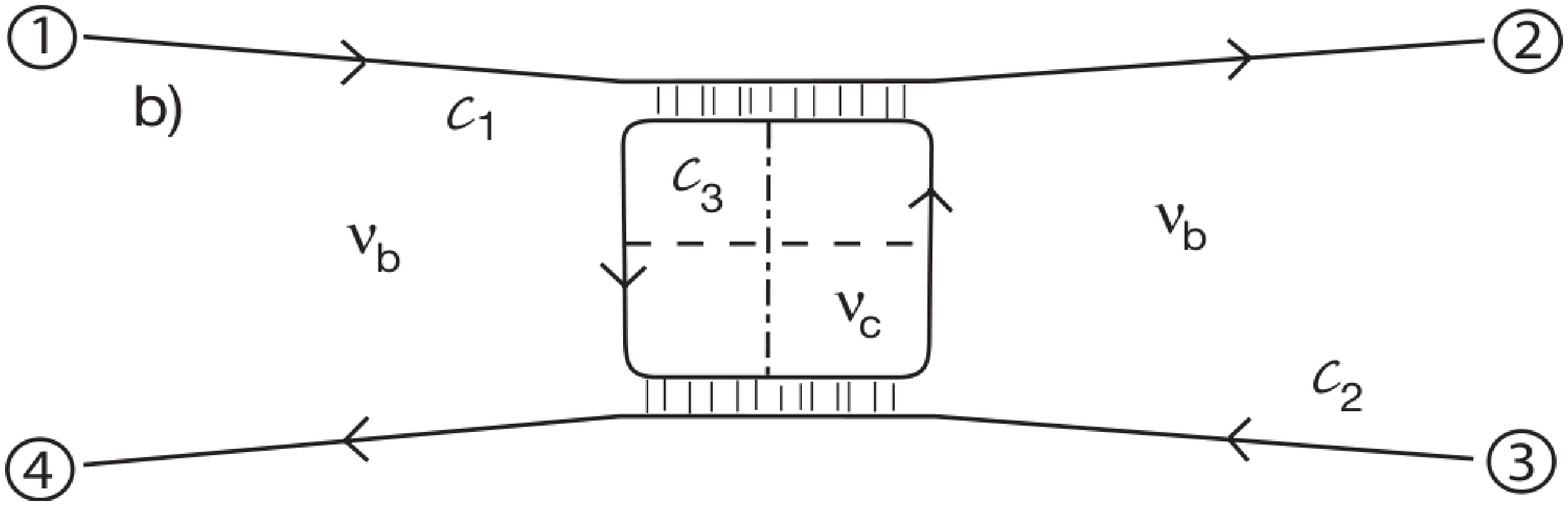}
\caption{Sketch of a Hall bar with a low density region in a constriction.
(a)  Simple edges bulk-vacuum and constriction-vacuum, composite edges
with two counter-propagating modes between bulk and constriction. 
In the lower panel (b), bulk-vacuum and bulk-constriction edges are simple, while  constriction-vacuum edges are composite.
Thin full lines represent intra-edge scattering, dashed and dashed-dotted lines
inter-edge scattering.
}
\label{setup.fig}
\end{figure}
%

If there is inter-edge scattering across the constriction, its strength is renormalized by the neutral mode dynamics. In the idealized model of 
a fully equilibrated constriction in the diffusive regime, we calculate the 
scaling dimensions of the most relevant operators for backwards and forward 
scattering. If the bare scattering matrix elements for both processes are comparable, the renormalization determines which one is dominant, and 
whether a zero bias peak or zero bias dip in the differential resistance is 
expected. For $\nu_b=1$,  our calculation agrees with the 
result of the particle-hole transformation  \cite{Roddaro+05,Lal07}, 
while we find a different result for some $\nu_b \neq 1$. We compare our predictions for zero bias peaks and dips for different 
 filling fractions with experiments.

 Interest in the physics of FQH neutral edge modes has been revived by the fact that the postulated non-Abelian statistics  of the $\nu=5/2$ FQH state is encoded 
in the dynamics of a neutral Majorana mode.~\cite{review} 
However, the $\nu = 5/2$ neutral mode also influences scattering across a 
constriction, and we find that for $\nu_b=3$ and $\nu_c=5/2$ the renormalization of scattering 
by the neutral mode may allow to distinguish the
 Pfaffian state 
from its particle-hole conjugate partner, the 
anti-Pfaffian.~\cite{Lee+07,LeHaRo07}.

{\em Description of low-density constriction:}
We model the 
setup  Fig.~1a  by two  chiral $\nu_b$ edges following paths ${\cal C}_1$ and 
${\cal C}_2$, 
and a closed $\nu_c$-edge surrounding the constriction region along a path ${\cal C}_3$. In the segments 
${\cal C}_1 \cap {\cal C}_3$ and ${\cal C}_2 \cap {\cal C}_3$ two 
edge  channels  are in spatial proximity to each other, and are 
 coupled both by a repulsive Coulomb interaction and scattering of charge 
$e$-electrons. The imaginary time Lagrangians for this setup  are 
%
\begin{eqnarray}
{\cal L}_0 & = & {1 \over 4 \pi \nu_b}  
\sum_{\alpha=1,2}  \int_{{\cal C}_\alpha} \! \! \! \! \! dx\ \partial_x \Phi_\alpha ( i \partial_\tau + v_b \partial_x)\Phi_\alpha  \nonumber \\
 & &  + {1 \over 4 \pi \nu_c}  \int_{{\cal C}_3} \! \! \! \! \! dx\ \partial_x \Phi_3 (
- i \partial_\tau + v_c \partial_x) \Phi_3 \label{Lzero.eq} \\
{\cal L}_{\rm int}  &= &  \sum_{\alpha=1,2}  v_{bc}  
 \int_{{\cal C}_\alpha \cap {\cal C}_3}  \! \! \! \! \! dx\ (\partial_x \Phi_\alpha)(\partial_x \Phi_3)\\
 {\cal L}_{\rm scat} & = &   \sum_{\alpha=1,2}
 \int_{{\cal C}_\alpha \cap {\cal C}_3}  \! \! \! \! \! dx \left[ \xi(x) e^{i \Phi_\alpha/\nu_b + \Phi_3/\nu_c} + c.c.~\right]  , \label{scattering.eq}
\end{eqnarray}
%
with ${\cal C}_3$ defined in the counterclockwise direction. 
Here, ${\cal L}_0$ describes simple edges propagating with velocities $v_b$ and $v_c$ along contours
${\cal C}_i$, $i=1,2,3$. For setup Fig.~1b, ${\cal C}_3$  is defined in the 
clockwise direction, and the dynamics  of $\Phi_3$ is governed by 
the  LL parameter $ \nu_b - \nu_c$.  
The electron density on edge $i$ is 
described by ${1 \over 2 \pi} \partial_x \Phi_i$.   Interaction and scattering of electrons between
the bulk and constriction edges are described by ${\cal L}_{\rm int}$ and 
${\cal L}_{\rm scat}$, respectively. $\xi(x)$ is a complex Gaussian random variable with mean zero and variance $\overline{ \xi^\star(x) \xi(x^\prime)}
= W_0 \delta(x - x^\prime)$. The disorder defines a bare mean free path 
$\ell_0 \sim 1/W_0$.

{\em Intra-edge scattering:} For  discussing  the influence of edge equilibration on transport properties, 
we discuss  the setup Fig.~2 and the specific choices $\nu_b = 1$,  
$\nu_c = 1/3$,  and generalize  to full constrictions as in  
Figs.~1a,b later. For this special choice of parameters, the interaction region between bulk and constriction edges is equivalent to the composite 
$\nu=2/3$ edge discussed in \cite{KaFiPo94}. In the following, we briefly summarize the main results of this reference. The disorder variance scales under an RG transformation as ${d W \over dl}  =  (3 - 2 \Delta) W $.
The scaling dimension $\Delta$  of the scattering operator 
Eq.~(\ref{scattering.eq}) flows under the RG as well, it has the initial value 
$ \Delta_0 = 2 \sqrt{3}(v_b + v_c - v_{bc})/  \sqrt{3(v_b + v_c)^2 - 4 v_{bc}^2}$.
For $\Delta_0 < 3/2$, scattering between the two edges is relevant, and 
the RG flows to a fixed point characterized by  $\Delta^\star  =1$ and 
disorder strength $W^\star$. At this random fixed point,  the composite edge 
formed in the interaction region between fields $\Phi_1$ and $\Phi_3$
is described by a charge mode $\Phi_\rho = \Phi_{1} + \Phi_3$ and a neutral mode $\Phi_\sigma = 
\Phi_{1} + 3 \Phi_3$, which are decoupled from each other. 
The fixed point Lagrangian  for the interacting region in Fig.~2 is 
%
\begin{eqnarray}
{\cal L^\star_{\rho \sigma}}  & = &  \int_0^L  \! \!  dx\ \left[ {3 \over 8 \pi }\partial_x \Phi_\rho (i \partial_\tau
+ v_\rho \partial_x)\Phi_\rho \right.
\label{chargedneutral.eq}
\\
& &\hspace*{-.5cm} +  \ \left.  {1 \over 8 \pi} \partial_x\Phi_\sigma(-i \partial_\tau + v_\sigma \partial_x)
\Phi_\sigma  +  \xi(x) e^{i \Phi_\sigma} + c.c.~\right]  .\nonumber
\end{eqnarray}
%

\begin{figure}[t]
\includegraphics[width=0.9\linewidth]{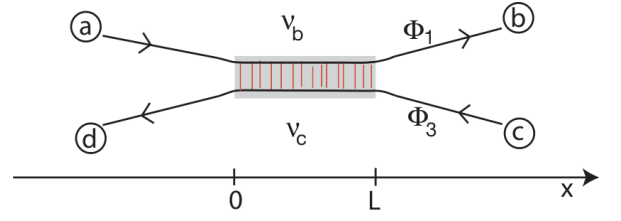}
\caption{Sketch of the composite edge between a $\nu_b$ and  $\nu_c$ incompressible quantum Hall regions. In the shaded contact region there 
is both interedge tunneling of electrons and a local Coulomb repulsion between 
counterpropagating edge modes.}
\label{setup.fig}
\end{figure}

{\em Transport - diffusive regime:}  To calculate  transport  through a composite $2/3$-edge of finite length $L$, we assume 
 that it  can be approximately described by the fixed point action 
Eq.~(\ref{chargedneutral.eq}). We define a conductance $G_{ab}=(\partial I_b/\partial V_a)|_{V_c}$, where
$I_b$ is the edge current at point $b$, and $V_a$, $V_c$ are voltages at 
contacts $a$ and $c$ (see Fig.~2). It 
  can be obtained from the Green function 
$g_{11}(x ,x_0; \omega_n)$ for field $\Phi_1$ via the Kubo formula
%
\begin{equation}
G_{ab} \ = \ {e^2 \over h 2 \pi } \ \lim_{\omega \to 0} v\;  \partial_x 
\left. g_{11}(x,x_0;\omega_n)\right|_{i \omega_n \to \omega  + i \eta} \ \ , 
\label{conductance.eq}
\end{equation}
%
where $x_0 < 0$ is in the vicinity of contact $a$ and $x>L$ in the vicinity of contact $b$. 
We derive a differential equation for the matrix Green function 
${\bf \sf g}(x,x_0;\omega_n)$
of fields $\Phi_1$, $\Phi_3$  from an  action analogous to Eq.~(\ref{Lzero.eq}) for $x<0$ and $x>L$ 
 and the fixed point action Eq.~(\ref{chargedneutral.eq}) for the 
region $0 < x <L$.  We  demand that ${\bf \sf g}$ is continuous everywhere, and 
decompose it into a particular part  with a discontinuity in its 
first derivative at $x=x_0$, and a homogeneous part, which is 
a solution of the differential  equations derived from extremizing the actions 
Eq.~(\ref{Lzero.eq}) and Eq.~(\ref{chargedneutral.eq}). 
In the diffusive regime,  
we find for the disorder averaged 
 $\Phi_1$-$\Phi_1$-Green function
%
\begin{equation}
\overline{g}_{11}(x,x_0; \omega_n) \ = \ 
{2\over 3} {2 \pi \over \omega_n} 
{e^{\omega_n{L + x_0 - x\over v_1} } \over g_\rho(0;\omega_n) - {1 \over 3} \overline{g}_\sigma(0;\omega_n)}  \ - \ {\pi \over \omega_n}
 \ \ .
 \label{oneoneaverage.eq}
\end{equation}
%
Here, $g_\rho$ satisfies the differential equation $(\omega_n + v_\rho \partial_x)
g_\rho =0$ subject to the boundary condition $g_\rho(L,\omega_n)=1$. In the 
differential equation for $g_\sigma$
%
\begin{equation}
\left( \omega_n + v_\sigma \partial_x - {v_\sigma \over \ell} \right) 
\overline{g}_\sigma(x;\omega_n) 
 \ = \ 0  \ , 
 \label{neutralgreen.eq}
\end{equation}
%
the nonlinear terms originating from  
Eq.(\ref{chargedneutral.eq}) are replaced by a self-energy term \cite{KaFiPo94},  
and  the boundary condition $\overline{g}_\sigma(L,\omega_n)=1$ is imposed.
 Combining the solution of Eq.~(\ref{neutralgreen.eq}) with Eqs.~(\ref{conductance.eq}),(\ref{oneoneaverage.eq}), we obtain for the average
conductance $\overline{G}_{ab} = 2/[3 - \exp(-L/\ell)]$, in agreement with  the result in \cite{SeAg08}.
Neglecting the renormalization of $\Delta$ in the range $1 < \Delta < 3/2$,  the temperature scaling of the equilibration length is 
%
\begin{equation}
\ell(T) \ = \ \ell_0 \left(T/T_{\rm gap}\right)^{2 - 2 \Delta} \ \ .
\label{temperature.eq}
\end{equation}
%
Here, $T_{\rm gap}$ is the high energy cutoff, which is on the order of the smaller of the two $\nu_b$, $\nu_c$ energy gaps.
In this simple picture, the equilibration length has the temperature dependence Eq.~(\ref{temperature.eq}) as long as $L_T < \ell(T)$. 
The renormalization of disorder stops when $L_T$  exceeds $\ell(T)$, giving rise to a zero temperature  decay length $\ell^\star = (v_\sigma /T_{\rm gap}) (\ell_0 T_{\rm gap}/v_\sigma )^{1/(3 - 2 \Delta)}$. 
 The temperature dependence of the  disorder averaged conductance $\overline{G}_{12}$ between contacts one and two
 can be obtained from a 
chiral network model for the setup Fig.~1a as
%
\begin{equation}
\overline{G}_{12} \ = \ 2  \ {e^2 \over h}\ {1 - e^{-L/\ell(T)} \over 3 +  e^{-L/\ell(T)}} \ \ 
\end{equation}
%
and decreases with decreasing temperature.  The conductance for 
the setup Fig.~1b is obtained in a similar way; it increases with decreasing temperature.

{\em Coherent transport:} in the regime of coherent transmission $L_T, L_V  > L$, the individual realization of disorder 
determines the conductance. 
To  calculate the distribution function 
of the  conductance we make use of the exact solution of the fixed 
point Lagrangian Eq.~(\ref{chargedneutral.eq}) 
\cite{KaFiPo94}. The neutral 
part can be mapped onto two co-propagating
free fermions $\Psi_1$, $\Psi_2$,  where the operator $\partial_x \Phi_\sigma$
corresponds to $\Psi^\dagger \sigma_z \Psi $. 
The random terms in Eq.~(\ref{chargedneutral.eq}) can be eliminated by 
transforming to new fields $\tilde{\Psi}(x) = U(x) \Psi(x)$, 
with the random SU(2) rotation $U(x)$ defined by
%
\begin{equation}
U(x) \ = \ P \ \exp\left[ - i \int_{x_0}^x dx \left( \xi(x) \sigma^+ + c.c. \right) \right] \ \ .
\end{equation}
%
Here, $P$ is the path ordering operator, and $\sigma^+ = \sigma_x + i \sigma_y$ is a linear combination of two Pauli matrices. 

To calculate the exact neutral Green function $g_\sigma(0,L;\omega_n)$, we 
assume that the random scattering takes place in the region $[0+\epsilon, L - \epsilon]$.
Then, the SU(2) rotation $U(x)$ has no position dependence in a 
neighborhood of 
$x=0$ and $x=L$, and we can integrate 
$\langle \partial_{x_1} \Phi_\sigma \partial_{x_2} \Phi_\sigma \rangle$, 
which is equal to the exactly known 
 $\langle  \Psi^\dagger(x_1) \sigma_z \Psi(x_1) \Psi^\dagger(x_2) \sigma_z \Psi(x_2)\rangle$,  with respect to $x_1$ and $x_2$ to obtain
 %
 \begin{equation}
 g_\sigma(0; \omega_n) \ = \  {\rm Tr}\left[ \sigma_z U(L)^\dagger \sigma_z U(L)\right]  e^{-L \omega_n \over v_\sigma}
  \ \ .
 \end{equation}
 %
 The trace of spin operators on the r.h.s.~is 
 equal to  the cosine of the angle $\Theta_1$ between the original spin quantization axis and 
 the rotated axis. Using this result in an equation for  ${g}_{11}(x,x_0; \omega_n)$ analogous to  Eq.~(\ref{oneoneaverage.eq}) but with $\overline{g}_\sigma $
 replaced by $ g_\sigma(0; \omega_n)$,   we find   %
 \begin{equation}
 G_{ab}(\cos \Theta_1) \ = \ {e^2 \over h} \ {2 \over 3 - \cos \Theta_1} \ \ .
 \end{equation}
 %
 For distances $L \gg \ell$,  the rotations $U(L)$ are uniformly distributed over the SU(2)-sphere, and
 $\cos \Theta_1$ is uniformly distributed in 
  $[-1,1]$. The minimum value $G_{ab}(-1)= e^2/(2h)$ 
 agrees with the minimum conductance found for a model with nonrandom scattering \cite{PoAv06}.

To calculate the conductance of the low-density constriction  Fig.~1a
in the coherent regime, we denote the SU(2)-angle of the 
left composite edge by 
 $\theta_1$ and the angle for the right composite edge by $\theta_2$.
  If interference contributions due to 
 paths winding around the $\nu_c$-region can be neglected,  the total conductance is
%
\begin{equation}
G_{12} \ = \ {e^2 \over h} \ {(1 - \cos \Theta_1)(1 - \cos \Theta_2) \over 
3 - \cos \Theta_1 -  \cos \Theta_2 - \cos \Theta_1 \cos \Theta_2} \ \ .
\label{Grandom.eq}
\end{equation}
%
 In the limit of adiabatic junctions 
\cite{CkHa98} with $\cos \Theta_1 = \cos \Theta_2 =-1$, the constriction  is fully transparent.  For either $\cos \theta_1=1$ or $\cos \Theta_2=1$ it is fully reflecting.

{\em Inter-edge scattering:} Besides the degree of 
composite edge equilibration,  transport through a low-density constriction can be 
influenced by backwards and forward scattering across the constriction 
(dashed and dash-dotted lines in Figs. 1a,b). The relative importance of forward and backwards scattering is controlled by both the bare tunneling matrix elements and LL renormalization. 
The bare matrix elements  depend in an exponential way on the tunneling distance and  are expected to be strongly influenced by the constriction geometry: for a long, "tunnel" like constriction, backwards scattering may be favored over forward scattering, and for a short constriction, forward scattering may be more important. For a small incompressible region, bare matrix elements may not depend too strongly on the constriction geometry, such that 
there is a parameter regime, for which renormalization and scaling dimensions 
may determine the most important tunneling process.  

We now consider an idealized model of an  infinitely long  and fully 
equilibrated constriction, and assume that the bare tunneling matrix elements for 
competing processes are of comparable size. 
We denote the edge creation operator for quasi-particles (QPs) 
by $\hat{T}(x,t)$ and define its local scaling dimension $g$ by $\langle \hat{T}^\dagger(x,t) \hat{T}(x,0)\rangle \sim t^{- g}$. 
The  voltage dependence of the  scattering probability involving QP operators 
with scaling dimension $g$ is $\sim V^{- 2(1-g)}$, hence  the process with the 
smallest scaling dimension is the most relevant one and will dominate 
transport  in the low energy limit.

On a simple $1/3$-edge,  e.g.~on the upper and lower edge in Fig.~1a and the left and right edge in Fig.~1b with $\nu_b=1$ and $\nu_c=1/3$, 
the operator for $e/3$ QPs  is unique and given by  $\hat{T}(x,t) = e^{- i \Phi_3(x,t)}$, and  its scaling dimension
is  $g_{\rm simple}=1/3$. At a position on the composite edge, e.g.~the left or right edge 
in Fig.~1a and top or bottom edge in Fig.~1b,  one has
 has to decompose $\Phi_3 = (\Phi_\sigma - \Phi_\rho)/2$ and has to evaluate the expectation value with respect to the Lagrangian 
Eq.~(\ref{chargedneutral.eq}). One finds $g_{\rm composite}=2/3$.
 There 
are two other operators with the same 
scaling dimension, creation of  charge $1/3$ QPs by $e^{i (\Phi_\rho + \Phi_\sigma)/2}$ and of charge $2/3$ QPs by $e^{i \Phi_\rho}$. 
Since $g_{\rm simple} < g_{\rm composite}$,  scattering between  
two simple edges is more relevant than  scattering between 
composite edges. Hence, as a function of source-drain voltage, one 
would expect a zero bias peak in the differential resistance for the setup Fig.~1a
and a zero bias dip for the setup Fig.~1b.  

In the same way, a bulk filling factor   $\nu_b=1$ and more general  constriction fillings from the hierarchy 
$\nu_c = n/(2 n \pm 1)$  \cite{Haldane83,Halperin84} with integer $n$ 
 can be analyzed. For  $\nu_c < 1/2$, 
 backscattering across the constriction 
 is more relevant than forward scattering (see Table I),   and causes
 a zero bias peak in the differential resistance. On the other hand,  for 
 $\nu_c > 1/2$  forward scattering is more relevant and gives rise to a zero bias dip in the differential resistance, in agreement with the experiments 
 \cite{Rodarro+04,Roddaro+05} and the particle-hole transformation argument 
\cite{Roddaro+05,Lal07}.
\begin{table}[t]
\begin{tabular}{|c|c|c|c|} \hline
$\nu_b$  & $\nu_c$ & $g_{\rm backwards}$ & $g_{\rm forward}$  \\ \hline
1 & $2 \over 3$ & $2 \over 3$ & $1 \over 3$ \\ \hline
1 & $1 \over 3$ & $1 \over 3$  & $2 \over 3$ \\ \hline
1 &  $ n \over  2 n \pm 1$ & $ n \over  2 n \pm 1$ & 
$ 1 - {n \over  2 n \pm 1}$ \\ \hline
$1 \over 3$ & $2 \over 7$ & $2 \over 7$ & $3 \over 7$ \\ \hline
$n \over 4 n -1$ & $ n + 1 \over 4 n + 3$ & $n + 1 \over 4 n +3$ & $4 n -1 \over 4 n + 3$ \\ \hline
\end{tabular}
\caption{Scaling dimension of the most relevant backwards and forward scattering operator for a constriction with filling fraction $\nu_c$ embedded in a  bulk with filling fraction $\nu_b$.}
\label{table0a.tab}
\end{table}

Can one expect a similar crossover between peak and dip in the differential 
resistance for $\nu_b < 1$ ? An obvious candidate to consider  is $\nu_b=1/3$ and $\nu_c=2/7$, 
as $\nu_c=2/7$ is an incompressible state of charge $-e/3$ holes and thus 
analogous to the $2/3$ state, which is an incompressible state of charge $-e$ holes.  The most relevant backwards scattering process between two random 
$\nu_c=2/7$ edges
has  scaling dimension $2/7$, whereas
the most relevant forward scattering process has scaling dimension $g = 3/7$, 
making backwards scattering
more relevant than forward scattering. 
From this argument, one would expect 
a zero bias peak in the differential resistance, which was 
not observed experimentally  \cite{Rodarro+04}.
This discrepancy 
may be due to 
the influence of the constriction geometry on  bare matrix elements or due to other nonuniversal effects \cite{RoHa02,PaMa04}. Such nonuniversal effects may also explain the crossover between dip and peak in the differential conductance as a function of gate-voltage
observed in \cite{chung+03}.

{\em Constriction filling $\nu_c=5/2$:} Charge transport through a low density constriction with  $\nu_b=3$ and $\nu_c = 5/2$ can  help to distinguish between two possible 
candidates for the 5/2-FQH state: the Pfaffian Pf and its particle hole 
conjugate, the anti-Pfaffian APf~\cite{Lee+07,LeHaRo07}, which   are topologically different 
from each other and differ in their edge structure.
The scaling dimension for scattering  of charge $e/4$ QPs between two Pfaffian edges is $g_{\rm Pf}=1/4$, whereas the scattering between 
two APf edges described by their random fixed point is $g_{\rm APf}=1/2$. 
As the edge between the Pf and $\nu_b=3$ is equivalent to edge between
APf and $\nu=2$,  in the idealized constriction model  the most relevant scattering process for a Pf state in the constriction region is backscattering, causing a  zero bias peak.   For an APf state in the constriction on the other hand, 
the most relevant scattering process is forward scattering, 
and a zero bias dip in the differential resistance is expected. If the dominant scattering process is determined by 
renormalization and not by bare matrix elements , the experiment by Miller et al.~\cite{Miller+07} is evidence for the Pf state to be the preferred ground state for filling fraction $5/2$. However, a recent experiment by Radu et al.~\cite{Radu08}  using samples with a filling fraction $5/2$ in both
the bulk and the constriction region is best described by a tunneling exponent  $g_{\rm APf}=1/2$.

In summary, we have discussed how the presence of composite 
edges around a low-density constriction influences transport in important ways.
 For a constriction in the diffusive regime, incomplete
equilibration of composite edges gives rise to a voltage and temperature 
dependence of the conductance. Even more strikingly, in the limit of coherent transport through composite edges, we predict  universal conductance oscillations and calculated their full distribution function.   
The LL renormalization of inter-edge scattering across the  constriction region determines whether the most relevant scattering process is forward or backwards scattering. 

{\em Acknowledgments.} 
Work was supported by NSF grant DMR 05-41988,
and by the Heisenberg program of DFG.


\end{document}